\documentclass[aps,pre,showpacs,twocolumn,superscriptaddress,floatfix]{revtex4}
\usepackage{graphicx}

\def \Im {\mbox{\cal Im}\,}

\begin{document}
\title{Traveling waves of infection in the Hantavirus epidemics}

\date{\today}

\author{G. Abramson}
\email{abramson@cab.cnea.gov.ar}
\affiliation{Center for Advanced Studies and Department of Physics
and Astronomy, University of New Mexico, Albuquerque, New Mexico 87131}
\affiliation{Centro At\'{o}mico Bariloche and CONICET,
8400 S. C. de Bariloche, Argentina}

\author{V. M. Kenkre}
\email{kenkre@unm.edu}
\affiliation{Center for Advanced Studies and Department of Physics
and Astronomy, University of New Mexico, Albuquerque, New Mexico 87131}

\author{T. L. Yates}
\affiliation{Department of Biology and Museum of Southwestern Biology,
University of New Mexico, Albuquerque, New Mexico 87131}

\author{R. R. Parmenter}
\affiliation{Sevilleta Long-term Ecological Research Program, Department of
Biology, University of New Mexico, Albuquerque, New Mexico 87131}

\begin{abstract}
Traveling waves are analyzed in a model of the Hantavirus infection of deer
mice. The existence of two kinds of wave phenomena is predicted. An
environmental parameter governs a transition between two regimes of
propagation. In one of them the front of infection lags behind at a constant
rate. In the other, fronts of susceptible and infected mice travel at the same
speed, separated by a constant delay. The dependence of the delay on system
parameters is analyzed numerically and through a piecewise linearization.
\end{abstract}

\pacs{87.19.Xx, 87.23.Cc, 05.45.-a}

\maketitle

\section{Introduction}

The Hantavirus zoonosis is a persistent problem in many regions of the world
\cite{mills99a}. Each species of Hantavirus is almost exclusively associated
with a single rodent species from which, eventually, humans may result
infected. The disease can range from mild to very severe, with a mortality in
excess of 50\%. Such is the case of the Hantavirus Pulmonary Syndrome (HPS)
produced by the virus associated with the deer mouse (\emph{Peromyscus
maniculatus}). In the North American Southwest, indeed, a serious outbreak of
this disease in 1993 led to the identification of the virus and its association
with the deer mouse \cite{parmenter99}. Since then, enormous effort and
resources have been devoted to the understanding of the ecology and the
epidemiology of the virus-mouse association, with the ultimate goal of being
able to predict and prevent the risk for the human population
\cite{mills99b,cparmenter98}.

As observed by Yates et al. \cite{yates02}, the role of the environment seems
to be determinant in the prevalence of the infection within the mouse
population, and the related incidence of the HPS. Its has been observed that
the disease can disappear completely from a local population during times of
adverse environmental conditions, only to reappear sporadically
\cite{mills99b,cparmenter99}. Besides, there are indications of focality of the
infection in ``refugia'' of the mouse population. Both phenomena are most
probably related, the refugia acting as reservoirs of the virus during times
when the infection has disappeared from most of the landscape. When
environmental conditions change, the infection spreads again from these
refugia.

In a recent work~\cite{abramson2002}, Abramson and Kenkre have shown that a
simple epidemic model is able to display qualitatively these behaviors, as the
result of a bifurcation of the equilibrium states of the system as controlled
by the carrying capacity of the medium. The purpose of the present paper is to
analyze the dynamics of simple traveling waves, as a model of the mechanisms in
which an epidemic wave of Hantavirus might propagate into a previously
uninfected region.

\section{Spatially extended model}

The model of Ref.~\cite{abramson2002} is a mean-field continuous model which
has been intentionally kept simple enough to facilitate the comprehension of
the basic mechanisms, and at the same time to incorporate as many known facts
about the ecology and epidemiology of the biological system as possible. The
reader is referred to~\cite{abramson2002} for a detailed discussion, which we
summarize here. It is known that the virus does not affect properties such as
the mortality of the mice, so that no appreciable difference in death rate, for
example, is to be expected between susceptible and infected mice. It is also
not transmitted to newborns, so that all mice are born susceptible. The
infection is transmitted from mouse to mouse through individual contacts,
presumably during fights. More general facts of the ecology of
\emph{Peromyscus} indicate that adults occasionally shift their home range to
nearby locations, in particular if these are vacant~\cite{stickel68,vessey87}.
This enables us to model the transport of mice as a diffusion process. Finally,
intra-species competition for resources indicate a saturated population growth,
which has been observed to be of a logistic form in the
laboratory~\cite{terman68}. Logistic growth is also a well established metaphor
of the dynamics of a self-limitating population~\cite{murray}.

For the sake of simplicity, assume further that the only population structure
is the division of the whole population into susceptible and infected mice,
denoted by $M_S$ and $M_I$ respectively. With these ingredients, the model is
described by the following equations:
\begin{eqnarray}
\frac{\partial M_S}{\partial t}&=&bM -cM_S -\frac{M_S M}{K(x,t)} -aM_S M_I+D\nabla^2
M_S, \label{dmsxdt}\\
\frac{\partial M_I}{\partial t}&=&-cM_I -\frac{M_I M}{K(x,t)} +aM_S M_I +D\nabla^2
M_S. \label{dmixdt}
\end{eqnarray}
Observe that the carrying capacity $K(x,t)$, containing the most direct effect
of the environment on the mouse population, is allowed, in our model, a spatial
and a temporal variation, to accommodate for a diversity of habitats and
temporal phenomena. The latter comprise the yearly variation due to
seasonality, but also sporadic fluctuations such as droughts and El Ni\~{n}o
effects.

The sum of the two equations~(\ref{dmsxdt})-(\ref{dmixdt}) reduces to a single
equation for the whole population:
\begin{equation}
\frac{\partial M}{\partial t}=(b-c)M\left(1-\frac{M}{(b-c)\,K}\right)+D\nabla^2 M.
\label{fisher}
\end{equation}
This is Fisher's equation, originally proposed as a deterministic model of the
spread of a favored gene in a population~\cite{fisher36}, and which eventually
became a standard model for a self regulated field in a diversity of situations
\cite{murray,abramson2001}.

\begin{figure}
\centering
\resizebox{\columnwidth}{!}{\includegraphics{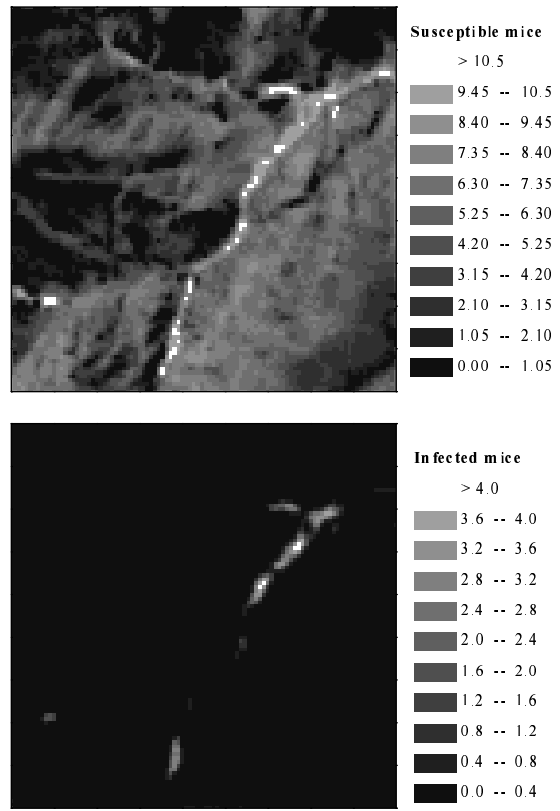}}
\caption{Density plots showing characteristic distribution of susceptible and
infected mice in an inhomogeneous landscape, where the carrying capacity has
been modeled according to the vegetation, derived from satellite imagery.}
\label{refugia2d}
\end{figure}

In Ref.~\cite{abramson2002} it was shown that, as a function of $K$, the system
undergoes a bifurcation between a stable state with only susceptible mice (and
$M_I=0$) to a stable state with both positive populations. The value of the
critical carrying capacity is a function of the parameters in the following
way:
\begin{equation}
K_c=\frac{b}{a(b-c)}.
\label{kc}
\end{equation}
This critical value does not depend on $D$, and the same bifurcation is
observed either in a space-independent system ($D=0$) or in a homogeneous
extended one in the presence of diffusion. In an inhomogeneous situation, for
moderate values of the diffusion coefficient, the infected subpopulation
remains restricted to those places where $K>K_c$, becoming extinct in the rest.
During times of adverse environmental conditions, these regions become isolated
and constitute the observed refugia. Figure~\ref{refugia2d} shows a typical
situation of this phenomenon. A simulated, albeit realistic, landscape of
$K(x)$ has been derived from satellite images in Northern Patagonia. The
carrying capacity is supposed proportional to the vegetation cover, and results
highest along a river, clearly inferred from the density plots. These show the
distribution of the populations of susceptible and infected mice. It can be
seen that susceptible mice cover most of the range. Meanwhile, the infected
population has become extinct in most of it, and persists only in some of the
places of highest $K$. The distributions shown in Fig.~\ref{refugia2d}
constitute a stable equilibrium of the system, found by numerical resolution of
Eqs.~(\ref{dmsxdt})-(\ref{dmixdt}) from an arbitrary initial condition, and
with zero-current boundary conditions.

\section{Traveling waves}

When conditions in the landscape change, how do the infected phase evolve from
the refugia, retracting from or, more importantly, invading previously
uninfected regions? this is the primary question we address in the present
paper. Fisher's equation (\ref{fisher}) has found wide applicability for the
description of the formation and propagation of spatio-temporal patterns.
Traveling wave solutions of Fisher's equation paradigmatically show how a
stable phase invades an unstable one in the form of a front propagating at a
certain speed.

There is no reason to suppose \emph{a priori} that the two waves, susceptible
and infective, will travel at the same speed. Accordingly, we use  an ansatz
which incorporates two independent traveling waves. In one dimension,
$z_1=x-v_St$ in Eq.~(\ref{dmsxdt}) and $z_2=x-v_It$ in Eq.~(\ref{dmixdt}). This
gives the following second-order system of ordinary differential equations:
\begin{eqnarray}
D\frac{d^2M_S(z_1)}{dz_1^2}+v_S\frac{dM_S(z_1)}{dz_1}+f(M_S,M_I)&=&0,\label{dmsdz}\\
D\frac{d^2M_I(z_2)}{dz_2^2}+v_I\frac{dM_I(z_2)}{dz_2}+g(M_S,M_I)&=&0,\label{dmidz}
\end{eqnarray}
where $v_S$ and $v_I$ are the speeds of the susceptible and infected waves
respectively, and $f$ and $g$ are the non-diffusive term in
(\ref{dmsxdt})-(\ref{dmixdt}).

\begin{figure}
\centering
\resizebox{\columnwidth}{!}{\includegraphics{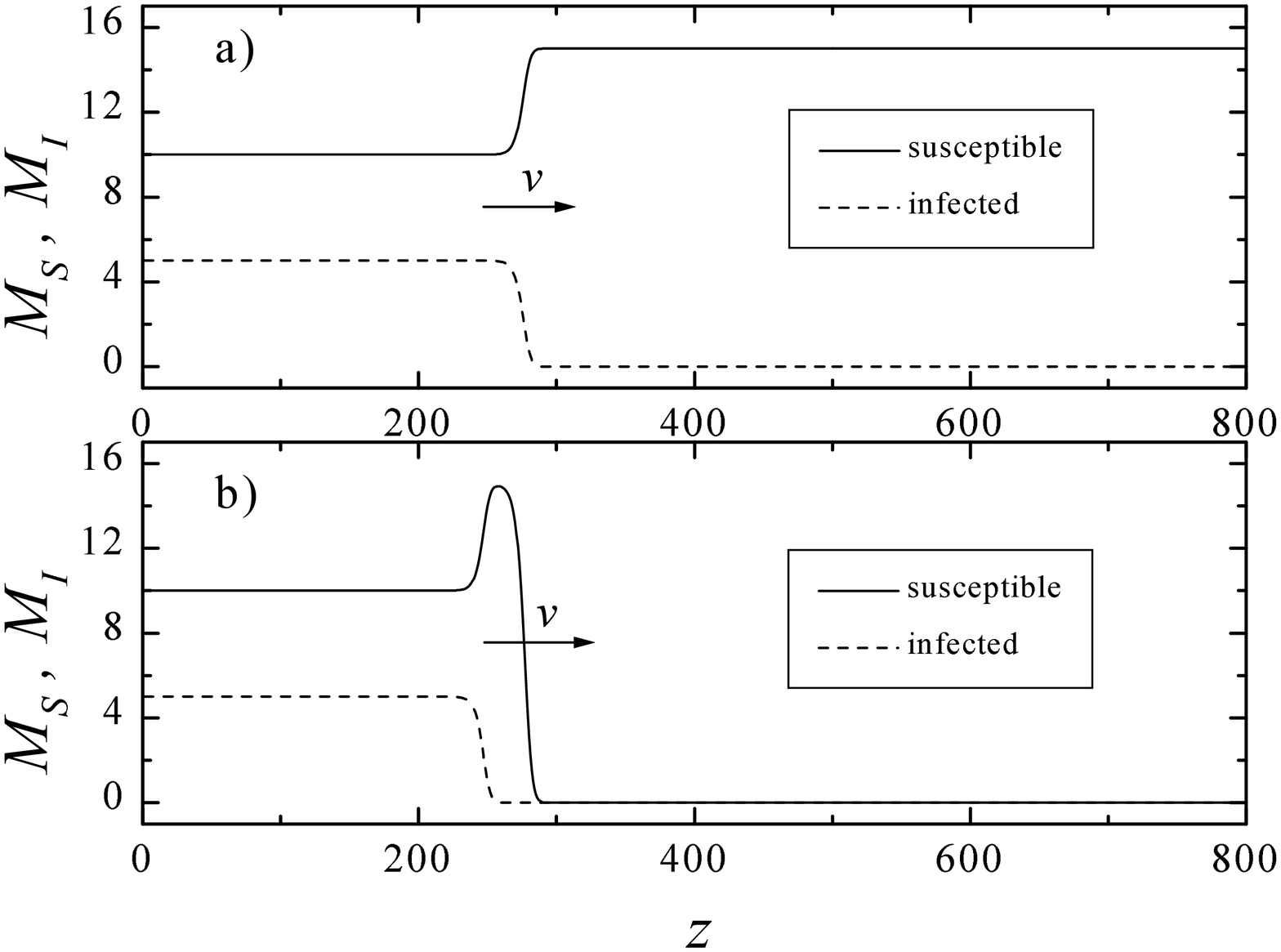}}
\caption{Traveling waves in the one dimensional model. a) Infection wave invading a
noninfected population. b) Noninfected mice invading an empty region, followed
by an infection wave. Model parameters: $a=0.1$, $b=1$, $c=0.5$, $D=1$, $K=30$.
Both waves move at the same speed $v=\sqrt{2}$ for this choice of parameters.}
\label{waves}
\end{figure}

There are two interesting scenarios for these waves. In the first one, a large
part of the system is initially at a state of low carrying capacity, below
$K_c$, and consequently the population consists of uninfected mice only, at the
stable equilibrium. Let us suppose that this region is in contact with a
refugium. If environmental changes occur, and the whole region finds itself at
a value of the carrying capacity $K>K_c$, the population will be out of
equilibrium. Two processes will occur simultaneously: the population of
susceptible mice will evolve towards a new equilibrium, and a wave of infected
mice will advance from the refugium, invading the susceptible population. The
speed of this wave can be calculated from the stability analysis of the
equilibrium states, requiring that the unstable infected mice density does not
oscillate below zero. This unstable equilibrium is $M_S^*=K(b-c)$, $M_I^*=0$,
and a linear stability analysis of the system (\ref{dmsdz})-(\ref{dmidz})
provides the following four eigenvalues:
\begin{eqnarray}
\lambda_{1,2}&=&\frac{-v\pm\sqrt{v^2+4D(b-c)}}{2D}, \\
\lambda_{3,4}&=&\frac{-v\pm\sqrt{v^2+4D[b-aK(b-c)]}}{2D}.
\label{lambda34}
\end{eqnarray}
The requirement that $M_I(z)$ does not oscillate below 0 imposes a restriction
to the radical in Eq.~(\ref{lambda34}), from which we find the following
expression for the speed of the traveling wave:
\begin{equation}
v\ge 2\sqrt{D\left[-b+aK(b-c)\right]}.
\label{vc1}
\end{equation}
An example of such a wave is shown in Fig.~\ref{waves}a, found by a numerical
integration of Eqs.~(\ref{dmsxdt}) and (\ref{dmixdt}) in 1 dimension.

The second interesting scenario corresponds to a system which is initially
empty of mice, both susceptible and infected. This situation is always unstable
within the validity of our simple model, but it is certainly a biological
possibility. Consider a system with $K>K_c$, and with $M_S=M_I=0$ in almost all
of its range, but in contact with a refugium in equilibrium. A wave of both
mice populations will develop invading the empty region. In fact, the total
population will display just the behavior of a traveling wave of Fisher's
equation. This wave will be composed of two fronts, susceptible and infected
respectively, with a delay of the latter with respect to the former. As before,
we can find the speeds of these two waves from a stability analysis around the
corresponding equilibria. The leading wave propagates into the null
equilibrium, $M_S^*=M_I^*=0$, to which the following eigenvalues correspond:
\begin{eqnarray}
\mu_{1,2}&=&\frac{-v_I\pm\sqrt{v_I^2+4Dc}}{2D},\\
\mu_{3,4}&=&\frac{-v_S\pm\sqrt{v_S^2-4D(b-c)}}{2D}.
\label{mu34}
\end{eqnarray}
In this situations, we require that $M_S(z)$ does not oscillate below 0, and
Eq.~(\ref{mu34}) provides the restriction on the speed of the susceptible
front:
\begin{equation}
v_S\ge 2\sqrt{D(b-c)},
\label{vs}
\end{equation}
which is, naturally, the same result as for Fisher's equation. The second
front, developed when part of the quasi-stable population of susceptible mice
is converted into infected, evolves from the equilibrium $M_S^*=K(b-c)$,
$M_I^*=0$, as in the previous scenario. Consequently, the same linear stability
analysis apply, and from Eq.~(\ref{lambda34}) we find a speed analogous to
Eq.~(\ref{vc1}). The front of infected move behind the susceptible one at a
speed:
\begin{equation}
v_I\ge 2\sqrt{D[-b+aK(b-c)]},
\label{vi}
\end{equation}
which, unlike $v_S$, does depend on the contagion rate $a$ and the carrying
capacity $K$. Figure~\ref{waves}b shows such a situation. The density of
susceptible mice rises from zero and lingers near the positive unstable
equilibrium before tending to the stable one. It is remarkable that a delay
exists between the two fronts, even when no such effect was explicitly
considered in the dynamics (such as incubation time, or age dependence). Such
delays have been observed in some populations and rationalized in different
ways (see~\cite{morgan00} or~\cite{mills99b} for a synthesis).

Even though Eqs.~(\ref{vc1}), (\ref{vs}) and (\ref{vi}) give only a lower bound
to the speed of propagation of the fronts, and allow a continuous of speeds
above this, in real situations only the lower bound is expected to be observed
as a stationary solution. Higher speeds may, however, play a role in transient
situations whose relevance in far from equilibrium systems such as real mice in
the wild, subject to a fluctuating environment, cannot be underestimated.

\begin{figure}
\centering
\resizebox{\columnwidth}{!}{\includegraphics{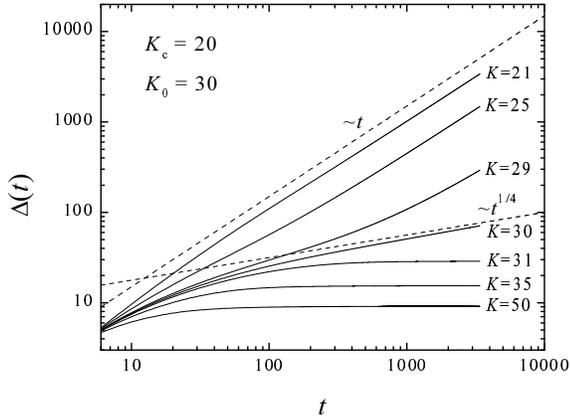}}
\caption{Delay of the infected front, as a function of time, following an initial
condition as described in the text. Two regimes are shown: $K_c<K<K_0$ and
$K_0<K$, separated by the critical case $K=K_0$, that behaves asymptotically as
$t^(1/4)$.}
\label{delay-vs-t}
\end{figure}

The different functional dependence of $v_S$ and $v_I$ on the parameters of the
system (Eqs.~(\ref{vs})-(\ref{vi})) indicates that two regimes are possible.
Indeed, when $v_I<v_S$ the front of infection lags behind the front of
susceptible at a delay $\Delta$ that increases linearly in time:
$\Delta(t)=(v_S-v_I)t$. Elementary manipulation of Eqs.~(\ref{vs}) and
(\ref{vi}) shows that this occurs whenever the carrying capacity satisfies:
\begin{equation}
K_c<K<K_0\equiv\frac{2b-c}{a(b-c)}
\label{k0}
\end{equation}
where $K_0$ is a new critical carrying capacity. At $K=K_0$ the delay becomes
effectively constant. For values of $K$ greater than $K_0$, the velocities
$v_I$ and $v_S$, calculated from linear considerations around the equilibria,
satisfy $v_I>v_S$. This regime is clearly unphysical in a stationary situation,
since the front of susceptible necessarily precedes the infected one. It could
be realizable and relevant in transient situations, that will be analyzed
elsewhere. From numerical resolution of the system, we can observe that
$v_I\rightarrow v_S$ and the delay tends to a constant value, whenever $K>K_0$.
Figure~\ref{delay-vs-t} shows the temporal evolution of the delay in the two
regimes $K<K_0$ and $K>K_0$, as well as in the critical case $K=K_0$, where it
is seen to increase as $\Delta\sim t^{1/4}$. It can be seen that there is a
transient time in both regimes, that gets progressively longer as $K$
approaches $K_0$ either from above or from below.

With the assumption that $v_S=v_I=v$, it is possible to perform a piecewise
linearization of Eqs.~(\ref{dmsdz})-(\ref{dmidz}) and find an approximate
analytical expression for the front shapes shown in Fig.~\ref{waves}b and,
consequently, for the delay $\Delta$ in the stationary state. The details of
the calculation can be found in the Appendix. The main result is the following
expression for the delay, for the case of equal speeds of the two classes of
mice:
\begin{equation}
\Delta=\frac{\sqrt{D}}{i\sqrt{(b-c)a(K-K_0)}}\log{(w_1 w_2)},
\label{delta1}
\end{equation}
where $w_1$ and $w_2$ are complex numbers of unit modulus that depend on $a$,
$b$ and $c$, so that the logarithm is effectively twice the phase difference
between them. When $K\rightarrow K_0^+$, the arguments of $w_1$ and $w_2$ tend
to $\pi$ and 0 respectively, so that the following critical behavior is
predicted by the linear approximation:
\begin{equation}
\Delta\sim \sqrt{D} [(b-c)a(K-K_0)]^{-1/2},\mbox{ when }K\rightarrow K_0^+.
\label{delta2}
\end{equation}

\begin{figure}
\centering
\resizebox{\columnwidth}{!}{\includegraphics{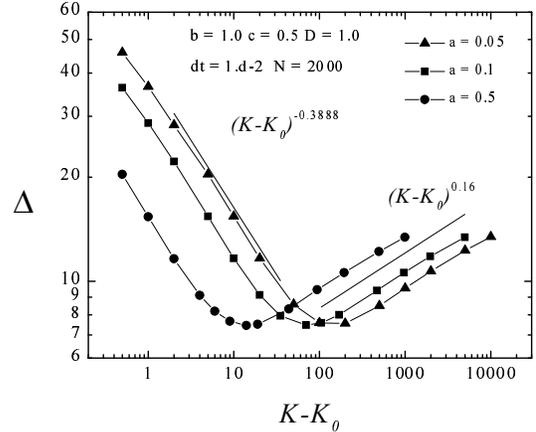}}
\caption{Delay of the infected front, as a function of $K$.}
\label{delay-vs-k}
\end{figure}

We have analyzed the dependence of $\Delta$ on $K$ by means of numerical
resolution of the full system. In Fig.~\ref{delay-vs-k} we plot the asymptotic
value of $\Delta$ as a function of $K-K_0$ for a variety of system parameters.
The behavior is found to be
\begin{equation}
\Delta\sim \sqrt{D} [(b-c)a(K-K_0)]^{-\alpha}
\label{delta3}
\end{equation}
for values of $K$ immediately above $K_0$. The exponent $\alpha$ is
parameter-independent, and its value, calculated from the simulations is,
approximately $0.388$. There is a discrepancy in the exponent found numerically
in the fully nonlinear system and the one found in the linearized
approximation, which shows the limitation of the linearized solution.

This critical behavior of the delay persists while the wavefronts are of the
kind shown in Fig.~\ref{waves}b. However, if $K$ continues to grow the
equilibrium value of the infected population turns greater than the equilibrium
value of the susceptible. There is a gradual crossover to a situation where
most of the population becomes infected. The value of the carrying capacity at
which this happens can be estimated from the known equilibria, $M_S^*=M_I^*$,
giving:
\begin{equation}
K_1=\frac{2b}{a(b-c)}.
\label{k1}
\end{equation}
Indeed, the numerical results show that the power law decay of the delay as a
function of $K$ starts to flatten for values of $K>K_1$ and reaches a minimum
which shows the same $a^{-1}$ dependence that $K_1$ does. Greater values of $K$
are probably unrealistic in the arid and semi-arid habitats of the Southwest.
It is nevertheless interesting to point out that the interaction of the two
fronts in this regime results in an \emph{increase} of the delay as a function
of $K$. This increase is also algebraic:
\begin{equation}
\Delta\sim(K-K_0)^\beta,
\label{delat2}
\end{equation}
with $\beta\approx 0.16$, as found in the numerical calculations.

\section{Conclusion}

We have analyzed the propagation of traveling fronts in 1 dimension in a simple
model of the ecology and epidemiology of the Hantavirus in deer mouse. We have
found that, when a mouse-free region is in contact with an infected region in
equilibrium, two waves propagate into the empty region. The first one is a wave
of susceptible mice. A wave of infected mice propagates behind it with a
certain delay. Two regimes of propagation exist, controlled by the
environmental parameter $K$. If $K_c<K<K_0$, the lag between the two fronts
increase linearly in time. Conversely, if $K>K_0$, the two fronts propagate at
the same speed and the delay depends critically on the difference $K-K_0$.

The occurrence of this double regime may be of relevance for the control of the
propagation of an epidemic wave. Indeed, the controlled reduction of $K$ ahead
of a propagating wave seems the most effective mean of stopping or reducing its
advance. Ideally, the carrying capacity should be reduced below $K_0$, to
ensure the complete extinction of the infection. However, if such a reduction
is not feasible, the fact that $K_0>K_c$ provides an alternative: a reduction
of the carrying capacity below $K_0$ would make the wave of infection start to
lag more and more behind the wave of healthy mice. Possible implementations of
these strategies, based on the propagation of waves in the presence of
``barriers,'' will be analyzed in detail elsewhere.

The existence of dynamical phenomena such as these traveling fronts also opens
the interesting possibility of subjecting our predictions to experimental
verification. Controlled experiments of front propagation could be possible in
the Sevilleta LTER facility, that the University of New Mexico operates near
Socorro, NM~\cite{lter}. Measurements of uncontrolled mice populations along
lines radiating from the refugia of infection will also provide evidence of the
propagation mechanisms. The observation of these in real mice population will
provide a valuable source of data to assign realistic values to the parameters
of the mathematical model.

\begin{acknowledgments}
V. M. K. and G. A. acknowledge many discussions with Fred Koster and Jorge
Salazar from which we learnt much regarding the peculiarities of the
Hantavirus. We also thank Greg Glass, Karl Johnson, Luca Giuggioli and Mar\'{\i}a
Alejandra Aguirre for discussions. V. M. K. acknowledges a contract from the
Los Alamos National Laboratory to the University of New Mexico and a grant from
the National Science Foundation's Division of Materials Research (DMR0097204).
G. A. thanks the support of the Consortium of the Americas for
Interdisciplinary Science and the hospitality of the University of New Mexico.
A part of the numerically intensive computations was carried out on the
Albuquerque High Performance Computing Center facilities.
\end{acknowledgments}

\section{Appendix: linearized solutions}

For the purpose of finding approximate solutions of the waves profiles, it is
better to replace the system (\ref{dmsdz})-(\ref{dmidz}) with one involving $M$ and $M_I$
instead:
\begin{eqnarray}
D M''+v M'+ (b-c)M-\frac{M^2}{K}&=&0, \label{dmdz}\\
D M_I''+v M_I'+q(M) M_I-a M_I^2&=&0, \label{dmidz-2}
\end{eqnarray}
where $q(M(z))=-c+a M(z)-M(z)/K$ and both speeds are assumed equal, as observed
in numerical resolutions for the regime $K>K_0$. Primes denote differentiation
with respect to $z$.  The reason for using this system is that the equation for
$M(z)$, Eq.~(\ref{dmdz}), being closed, can be solved independently of $M_I$.
Its solution can be used then into Eq.~(\ref{dmidz-2}) as a $z$-dependent
parameter and solve for $M_I(z)$.

\begin{figure}
\centering
\resizebox{\columnwidth}{!}{\includegraphics{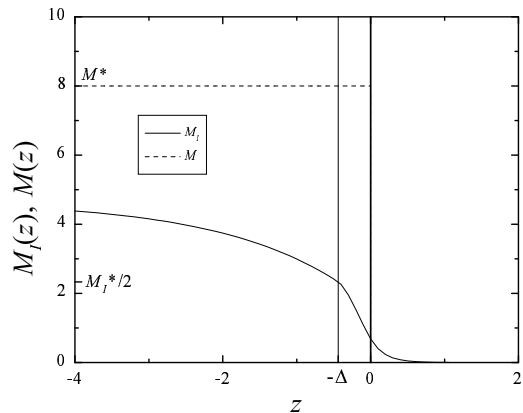}}
\caption{Linearized solutions. System parameters are: $a=0.3$, $b=1$, $c=0.5$, $K=16$, $D=0.1$.}
\label{wavefront}
\end{figure}

Linearized solutions for the traveling waves of Fisher's equation~(\ref{dmdz})
are well known, and essentially consist of two exponentials, matched smoothly
at $z=0$, representing a front that travels to the right at a speed $v\ge
2\sqrt{D(b-c)}$. Such a function needs a further simplification in order to
solve Eq.~(\ref{dmidz-2}). We approximate $M(z)$ with a step function
discontinuous at $z=0$: $M(z)=M^*=K(b-c)$ if $z<0$ and $M(z)=0$ if $z>0$.
Consequently, we have that $q(z)=a M_I^*=-b+aK(b-c)$ if $z<0$ and $q(z)=-c$ if
$z>0$. We also linearize on both sides of the center value
$M_I(-\Delta)=M_I^*/2$, which defines $\Delta$ as the delay between the two
waves. Figure~\ref{wavefront} shows the geometry of the procedure. The
linearized equation for $M_I$ breaks into three regimes:
\begin{eqnarray}
D M_I''+v M_I'- a (M_I-M_I^*)&=&0 \mbox{ if }z<-\Delta, \\
D M_I''+v M_I'+a M_I^* M_I&=&0 \mbox{ if }z\in[-\Delta,0], \\
D M_I''+v M_I'-C M_I&=&0 \mbox{ if }0<z.
\end{eqnarray}

The solutions of these are respectively:
\begin{eqnarray}
M_1(z)&=&-\frac{b}{a}+K(b-c)+a_1 e^{\lambda z} \mbox{ if }z<-\Delta, \label{m1}\\
M_2(z)&=&b_1 e^{\mu_+ z}+b_2 e^{\mu_- z}~~~~~~~~~~~\mbox{if }z\in[-\Delta,0], \\
M_3(z)&=&c_1 e^{\nu z}~~~~~~~~~~~~~~~~~~~~~~~~~\mbox{if }0<z, \label{m3}
\end{eqnarray}
where:
\begin{eqnarray}
\lambda&=&\frac{\sqrt{b-c+a}-\sqrt{b-c}}{\sqrt{D}}, \label{lambda}\\
\mu_\pm&=&-\sqrt{\frac{b-c}{D}}\pm i\sqrt{\frac{b-c}{D}a(K-K_0)},\\
\nu&=&-\frac{\sqrt{b-c}+\sqrt{b}}{\sqrt{D}}, \label{nu}
\end{eqnarray}
and the terms that diverge in infinity have already been cancelled. It is clear
that $\lambda>0$ and $\nu<0$, as required for the solution not to diverge in
$\mp\infty$. The real part of $\mu_\pm$ is always positive, and there is an
imaginary part when $K>K_0$.

The solutions $M_1$, $M_2$ and $M_3$ need to be matched smoothly at $z=-\Delta$ and
$z=0$, in the following way ($r=b-c$ from now on):
\begin{eqnarray}
M_1(-\Delta)&=&-\frac{b}{2a}+\frac{kr}{2},\\
M_2(-\Delta)&=&-\frac{b}{2a}+\frac{kr}{2},\\
M_1'(-\Delta)&=&M_2'(-\Delta),\\
M_2(0)&=&M_3(0),\\
M_2'(0)&=&M_3'(0).
\end{eqnarray}

These are five equations with five unknowns that can be solved without
difficulty:
\begin{eqnarray}
a_1&=&-\frac{-b/a+kr}{2}e^{\lambda \Delta}, \label{a1}\\
b_1&=&-\frac{(\lambda+\mu_-)}{2(\mu_+-\mu_-)}(kr-b/a)e^{\mu_+\Delta}, \\
b_2&=&-\frac{(\lambda+\mu_+)}{2(\mu_+-\mu_-)}(kr-b/a)e^{\mu_-\Delta}, \\
c_1&=&b_1+b_2,
\end{eqnarray}
and:
\begin{equation}
\Delta=\frac{1}{\mu_+-\mu_-}
\log\left[\frac{(\lambda+\mu_+)(\mu_--\nu)}{(\lambda+\mu_-)(\mu_+-\nu)}\right].
\label{delta4}
\end{equation}

Expressions (\ref{m1})-(\ref{m3}), (\ref{lambda})-(\ref{nu}) and
(\ref{a1})-(\ref{delta4}) define the linearized solution for $M_I$. An example
of such a front wave is shown in Fig.~\ref{wavefront}.

Using expressions (\ref{lambda})-(\ref{nu}) in (\ref{delta4}), it easy to see
find:
\begin{equation}
\Delta=\frac{\sqrt{D}}{i\sqrt{(b-c)a(K-K_0)}}\log{(w_1 w_2)},
\label{delta5}
\end{equation}
where
\begin{eqnarray}
w_1&=&\frac{\lambda +\mu_+}{\lambda +\mu_-} \equiv e^{i\phi_1},\\
w_2&=&\frac{\mu_--\nu}{\mu_+-\nu}\equiv e^{-i\phi_2},
\end{eqnarray}
where $\phi_1$ and $\phi_2$ are respectively the arguments of the complex
numbers $\lambda +\mu_+$ and $\mu_+-\nu$. It is easy to see that $\phi_1$ and
$\phi_2$ do not depend on $D$, so that the logarithm only corrects the
dependence of $\Delta$ on $(b-c)a(K-K_0)$. It can be also easily observed that
$w_1$ lies in the second quadrant, and $w_2$ in the fourth, and that
$\Im{w_1}=-\Im{w_2}$. When $K\rightarrow K_0^+$, the imaginary parts tend to
zero and the phase difference between both tend to $\pi$. Consequently the
leading behavior of $\Delta$ results:
\begin{equation}
\Delta\sim \sqrt{D} [(b-c)a(K-K_0)]^{-1/2}.
\label{delta6}
\end{equation}
This result from the piecewise linearization does not agree with that found
numerically in the nonlinear system.

\end{document}